\documentclass[prl,preprint,showpacs,preprintnumbers,amsmath,amssymb]{revtex4}

\usepackage{graphicx}
\usepackage{dcolumn}
\usepackage{bm}

\def\bfo{\begin{eqnarray*} }
\def\efo{\end{eqnarray*} }
\def\ba{\begin{eqnarray*} }
\def\ea{\end{eqnarray*} }
\def\beq{\begin{equation}}
\def\eeq{\end{equation}}

\def\det {\hbox{det}}

\def\p{\partial}

\def\p{\partial}

\def\ttp{\overline {\overline \rho}}
\def\det{{\text det}}

\begin{document}

\preprint{}

\title{Comment on ``Scattering Theory Derivation of a 3D Acoustic Cloaking Shell"}

\author{Allan Greenleaf\,${}^*$}
\affiliation{Department Mathematics
 University of Rochester, Rochester, NY 14627}

\author{Yaroslav Kurylev}
\affiliation{Department of Mathematical Sciences, University College London, Gower Street, London,
WC1E 6BT, UK}

\author{Matti Lassas}
\affiliation{Institute of Mathematics, Helsinki University of Technology, FIN-02015, Finland}

\author{Gunther Uhlmann}
\affiliation{$\hbox{Department of Mathematics, University of Washington, Seattle, WA 98195}$
}

\date{January 21, 2008}


\maketitle

In a recent Letter, Cummer et al. \cite{Cummer} give a description of 
material parameters for acoustic wave propagation giving rise to a 3D spherical cloak, and verify
the cloaking phenomenon on the level of scattering coefficients. A similar configuration has been
given in \cite{ChenChan}. In this Comment, we show that these theoretical constructions follow
directly from earlier work \cite[Sec. 3]{GKLU} on full wave analysis of cloaking for the
Helmholtz equation with respect to  Riemannian metrics. Furthermore, the analysis there covers the case of
acoustically radiating objects being enclosed in the cloaked region.

For a Riemannian metric $g=(g_{ij})$ in $n$-dimensional space, 
the Helmholtz equation with source term is 
\beq\label{one}
\frac1{\sqrt{|g|}} \sum_{i,j=1}^n \frac{\p}{\p x_i} \left(\sqrt{|g|} \, 
g^{ij} \frac{\p u}{\p x_j}\right) + k^2 u = f,
\eeq
where $|g|=\det(g_{ij})$ and $(g^{ij})=g^{-1}=(g_{ij})^{-1}$.
The equation of acoustics, in the notation of \cite{Cummer}, is
\beq\label{two}
\nabla\cdot(\ttp^{-1} \nabla p) +\frac{\omega^2}\lambda p=0.
\eeq
The connection between (\ref{one}) and (\ref{two}) is given by 
\beq\label{three}
k=\omega,\quad \ttp^{-1}=(\sqrt{|g|}\, g^{ij}), \quad \sqrt{|g|}=
\lambda^{-1}.
\eeq
In \cite{GKLU},  we showed that the  degenerate cloaking metrics $g$ 
for electrostatics constructed in \cite{GLU1}, 
giving the same boundary measurements as the Euclidian metric $g_0=(\delta_{ij})$, also
cloak with respect to solutions of the Helmholtz equation at any
nonzero frequency $k$ and with any source $f$. An example in 3D, with respect to
spherical coordinates $(r,\theta,\phi)$,  is
\beq\label{four}
g^{-1}=
\left(\begin{array}{ccc}
2(r-1)^2\sin \theta & 0 & 0\\
0 & 2 \sin \theta & 0 \\
0 & 0 &  2 (\sin \theta)^{-1}\\
\end{array}
\right)
\eeq
on $\{1<r\le2\}$, with the cloaked region being the ball $\{0\le r<1\}$. Note that  $\frac{\p}{\p\theta}, \frac{\p}{\p\phi}$ are not normalized to have length 1; otherwise, (\ref{four}) agrees with \cite[(24-25)]{Cummer} and \cite[(8)]{ChenChan}. This $g$ is the image
of $g_0$ under the singular transformation $(r,\theta,\phi)=F(r',\theta',\phi')$ defined by
$r=1+\frac{r'}2,\,\theta=\theta',\, \phi=\phi',\, 0<r'\le 2$,  which blows up the point
$r'=0$ to the cloaking surface $\Sigma=\{r=1\}$. The same transformation  was later used in
\cite{PSS1} and gives rise to the cloaking structure that is referred to in \cite{GKLU} as the {\emph{single coating}}. It was shown in \cite[Thm.1]{GKLU} that if the cloaked region is given
any nondegenerate metric, then finite energy waves $u$ that satisfy the Helmholtz equation (\ref{one}) on
$\{r<2\}$ in the sense of distributions have the same set of Cauchy data at $r=2$, i.e., the same acoustic boundary measurements,  as do the solutions for 
the Helmholtz equation for $g_0$ with source term $f\circ F$. (This also holds for Maxwell's
equations, explaining the ``mirage effect" of \cite{Guenneau}.) The part of $f$ supported within the cloaked region is undetectable at $r=2$. Furthermore, on the boundary
$\Sigma$ of the cloaked region, the normal derivative of
$u$ must vanish, so that within $\Sigma$ the acoustic waves propagate as if $\Sigma$
were lined with a sound--hard surface. 

\noindent{\bf Acknowledgements:}AG and GU
are supported by  NSF, ML by Academy of Finland, and GU by a Walker 
Family Endowed Professorship.

\noindent${}^*$Authors are listed in alphabetical order. Correspondence should be addressed to: \linebreak allan@math.rochester.edu.

\end{document}